\documentclass[a4paper]{article}

\usepackage[a4paper,top=3cm,bottom=2cm,left=3cm,right=3cm,marginparwidth=1.75cm]{geometry}

\usepackage{lineno}
\usepackage[dvipsnames]{xcolor}
\modulolinenumbers[5]

\usepackage{pstool}
\usepackage{setspace}
\usepackage{graphicx, tabularx}
\usepackage{float}

\usepackage[numbers,sort&compress]{natbib}
\usepackage[colorlinks=true, allcolors=blue]{hyperref}

\usepackage{nomencl}
\makenomenclature
\usepackage{etoolbox}
\renewcommand\nomgroup[1]{%
  \item[\bfseries
  \ifstrequal{#1}{P}{Parameters}{%
  \ifstrequal{#1}{N}{Non-dimensional numbers}{%
  \ifstrequal{#1}{A}{Abbreviations}{}}}]}
\newcommand{\nomunit}[1]{%
\renewcommand{\nomentryend}{\hspace*{\fill}#1}}

\usepackage{caption}
\usepackage{subcaption}
\usepackage{xcolor}

\usepackage{lscape}
\usepackage{booktabs}

\usepackage{multirow}
\usepackage{longtable}

\usepackage{amsmath}
\usepackage{textcomp}

\usepackage{ulem}
\usepackage{url}
\def\deg{$^\circ$C}
\def\We{\mathrm{We}}
\def\Re{\mathrm{Re}}
\def\Oh{\mathrm{Oh}}
\def\Pas{Pa\cdot s}
\def\Al{Al$_2$O$_3$} 

\title{Spreading-splashing transition of nanofluid droplets on a smooth flat surface}
\author{Y. T.~Aksoy$^a,\ast$, P.~Eneren$^a$, E.~Koos$^{b}$, M. R.~Vetrano$^a$}
 \date{ \small
 $^a$ {KU Leuven, Department of Mechanical Engineering, Division of Applied Mechanics and Energy Conversion (TME), B-3001 Leuven, Belgium} \\
$^b$ {KU Leuven, Department of Chemical Engineering, Soft Matter, Rheology and Technology (SMaRT), B-3001 Leuven, Belgium} \\
 $^{\ast}$ Corresponding Author \\
\textit{Email addresses}: yunus.aksoy@kuleuven.be (Y. T. Aksoy), pinar.eneren@kuleuven.be
(P. Eneren), erin.koos@kuleuven.be (E. Koos), rosaria.vetrano@kuleuven.be (M. R. Vetrano) \\}

\begin{document}

\maketitle

\begin{abstract}
\textit{Hypothesis}: Even a small fraction of nanoparticles in fluids affects the splashing behavior of a droplet upon impact on a smooth surface.\\
\textit{Experiments}: Nanofluid drop impact onto a smooth sapphire substrate is experimentally investigated over wide ranges of Reynolds ($10^2<\Re<10^4$) and Weber ($50<\We<500$) numbers for three nanofluid mass concentrations (0.01\%, 0.1\%, 1\%) using high-speed photography. Nanofluids are prepared by diluting a commercial \Al-water nanofluid in aqueous glycerol solutions without dispersants. In total, 30 samples are prepared and 1799 data points are acquired. Every sample is experimentally characterized prior to droplet impact measurements in terms of stability, density, viscosity, and surface tension to demonstrate the observed outcomes on the We-Re maps. Each droplet impact condition is repeated at least 3 times to ensure good repeatability.\\
\textit{Findings}: The non-monotonic behavior of the spreading-to-splashing transition remains the same for nanofluids. However, nanofluids influence this boundary by promoting splashing at low Reynolds numbers. We explain this behavior by increased lamella spreading speed and lift during the lamella spreading stage. Finally, we develop an empirical correlation which describes the splashing threshold dependency on nanoparticle concentration for the first time.. \\

\noindent\textbf{Highlights}
\begin{itemize}
    \item The We-Re maps are plotted for various nanoparticle concentrations.
    \item Nanofluids promote splashing for low $\Re$ numbers.
    \item Increased lamella spreading velocity due to nanoparticles leads to splashing.
    \item A generalized splashing threshold correlation incorporating the particle concentration is proposed and verified.
\end{itemize}

\end{abstract}

\section{Introduction}\label{sec:introduction}
Droplet impact on a solid surface has received considerable interest from both academical and industrial researchers as there is a wide variety of potential applications, such as in internal combustion engines (fuel droplets), inkjet printing, and spray cooling~\cite{Marengo2011, Yarin2006}. While there has been extensive literature into splashing of liquid drops, research into systems with added colloids, such as nanofluids, is still limited~\cite{Aksoy2021}. Meanwhile, the spreading-to-splashing transition of particle laden fluids plays a crucial role in many applications, ranging from inkjet printing in microelectronics~\cite{Minemawari2011} to forensic science~\cite{Laan2014, Josserand2016}.

Smooth deposition (spreading), complete bounce as a single drop, or splashing with secondary droplets are the possible outcomes for a droplet hitting a dry solid surface~\cite{Driscoll2010}. Splashing of a drop is defined as the disintegration of the impacting drop into two or more secondary droplets after a collision with a solid surface~\cite{Rein1993}. This splashing can take place in two different ways: prompt or corona. In prompt splashing, smaller droplets are released immediately after a high velocity impact via the breakup of the lamella tip, generally parallel to the substrate, without generating any thin-sheet or corona~\cite{Vega2017}. On the other hand, for corona splashing, a thin liquid lamella forms at the boundary of a droplet impacting at lower velocities. The lamella then rises from the substrate and the droplets formed by its breakup are ejected at a certain angle with respect to the surface~\cite{Josserand2016}.

The transitions between spreading, bouncing and splashing in liquid droplets are governed by the fluid density $\rho$, the surface tension $\sigma$, the dynamic viscosity $\eta$, the droplet impact speed $u_0$, the droplet diameter shortly before the impact $d_0$, the surrounding gas pressure, and the average surface roughness $R_a$ and wettability of the substrate~\cite{Yarin2006, Rioboo2001, Xu2007, Quetzeri2019}. These parameters can be incorporated into several non-dimensional groups~\cite{Zhang2017, Mundo1995}:
\begin{itemize}
    \item the Weber number $\We = \rho d_0 u_0^2 / \sigma$, which expresses the balance between the inertial and surface tension forces;
    \item the Reynolds number $\Re = \rho u_0 d_0 / \eta$, representing the ratio of inertial to viscous forces;
    \item the Ohnesorge number $\Oh = \We^{1/2}/\Re$, which relates the viscous forces to inertial and surface tension forces; and
    \item the surface roughness number $\textup{S} = R_a/d_0$, which is a measure of the relative surface roughness of the substrate.
\end{itemize}
Various semi-empirical correlations for the prediction of the splashing transition have been proposed. Generally, the splashing threshold is characterized by the splashing parameter $K = \We \cdot \Re^\alpha$, which incorporates the inertia, viscous stress, and fluid surface tension~\cite{Stow1981, Mundo1995}. The sign of $\alpha$ determines the viscosity effect on splashing. A negative $\alpha$ indicates that a lower viscosity prevents splashing whereas a positive sign implies that a lower viscosity promotes splashing. As both signs have appeared in literature, a two-piece function is derived and first proposed by Palacios et al. to include these signs depending on the viscosity values~\cite{Palacios2013}:
\begin{equation}
    \We_\mathrm{crit.} = K_1 \Re^{\beta_1} + K_2 \Re^{-\beta_2},
    \label{eq:splash}
\end{equation}
where $\We_\mathrm{crit.}$ represents the critical $\We$ number for the splashing threshold. The correlation defined in Eqn.~\ref{eq:splash} thus describes the splashing threshold in both low and high viscosity regions. In other words, the positive exponent of $\Re$, $\beta_1$, with the coefficient $K_1$, determines the splashing behavior at low viscosity values whereas the second term with coefficient $K_2$ dominates in the high viscosity region via the negative exponent $-\beta_2$. These limits and the full relation are shown in Fig.~S1.

Even if the aforementioned empirical formulations are valid tools for the splashing predictions, they do not provide a physical insight on the fundamental phenomena triggering the splashing occurrence. The physics behind this phenomenon is much more complex and remains under some debate. Mandre and Brenner~\cite{Mandre2012} report the interactions between the droplet and the substrate consisting of different stages. More specifically, the droplet starts to decelerate while approaching to the substrate, due to the compression of the air beneath. As a consequence, a stagnation point near the center develops and the liquid around it is diverted towards the periphery, forming a liquid sheet that continues to spread on the surface without touching it. If the lamella touches the substrate, a viscous boundary layer develops with a consequent deceleration of the lamella tip and its lift-off, thus resulting in splashing. However, an unstable levitated lamella alone does not necessarily splash. According to Thoroddsen et al.~\cite{Thoroddsen2010}, an instability in the levitated lamella may cause its tip to touch the solid surface ahead of the contact line, thus rupturing the levitated lamella by entrapping gas bubbles when these localized contacts meet the expanding contact line. This produces a powerful capillary-driven splash and multiple rings of entrapped micro-bubbles are observed in the works of Driscoll et al.~\cite{Driscoll2010} and Palacios et al.~\cite{Palacios2012}.

Indeed, for many years, the event of splashing is explained by invoking the role of Rayleigh-Taylor (R-T) instabilities~\cite{Allen1975}. Lately, the effect of the gas pressure on the splashing has implied that the phenomenon is instead driven by the appearance of Kelvin-Helmholtz (K-H) instabilities in the lamella according to Liu et al.~\cite{Liu2015}. They describe how the ultra thin air layer initiates the K-H instabilities around the lamella tip, inducing splashing. In particular, the lamella is separated from the substrate by an ultra thin air layer with a thickness down to 10~nm, i.e., high Knudsen number occurs due to this gap being smaller than the mean free path of air molecules~\cite{Kolinski2012}. Therefore, the interior airflow transfers momentum at a high velocity, which is comparable to the speed of sound, and it generates a stress 10 times greater than what is expected in continuum case. Hence, the splashing eventuates due to the stress generated by this ultra thin gas layer.

Since the ultra thin air layer influences splashing, the properties of the ambient gas play a significant role. For instance, the effect of the surrounding gas composition on the lamella is investigated by Burzynski et al.~\cite{Burzynski2019}. They observe an important impact of the gas density, gas viscosity, and mean free path of the gas molecules on the lift of the lamella and hence on the splashing occurrence. In fact, the ambient pressure may completely suppress the splashing in the case of its absence. Moreover, the threshold pressure can be incorporated as a function of droplet impact speed~\cite{Xu2005}. Besides, Riboux and Gordillo~\cite{Riboux2014} propose that lamella lift force has two main components: the lubrication force at the wedge formed between the substrate and the lamella tip; and the suction force on top of the drop. Their experiments compare well with the data of several others~\cite{Xu2005, Palacios2013, Stevens2014}.

Surface roughness is another important factor affecting the splashing threshold. Roisman et. al.~\cite{Roisman2015} observe that the main surface parameter is the characteristic slope of the substrate morphology. However, the effect of the surface roughness is only observed for prompt splashing, since corona splashing is mostly driven by the impact of the ambient gas conditions on the lamella lift~\cite{Xu2007}. In fact, the surface roughness, inhibiting thin-sheet formation, triggers the prompt splashing at the advancing contact line. On the other hand, lowered gas pressure suppresses the droplet ejection not only for the corona, but also for the prompt splashing~\cite{Latka2012}.

The studies and the relations presented above for splashing, however, are developed only for pure fluids or fluid solutions. In the last decade, nanofluids are proposed for a large number of applications, such as for impinging jets~\cite{Ersayin2013, Basaran2013, Buonomo2019} and sprays~\cite{Figueiredo2020, Aksoy2021, Kang2019, Maly2019}, owing to their enhanced thermal properties. Nevertheless, no investigation has been performed to evaluate the impact of the nanofluids, and in particular of the nanoparticle load in the bulk fluid, on splashing occurrence. Indeed, the nanoparticles have the potential to strongly modify the lamella formation. Therefore, in this framework, we experimentally investigate how the nanoparticle concentration in water- and glycerol-based \Al~nanofluids affects the splashing threshold. Consequently, we examine the spreading-to-splashing transition for a range of $\We$ and $\Re$ numbers from 50 to 500 and from $10^2$ to $10^4$, respectively. Finally, we propose a new correlation that incorporates the nanofluid mass fraction.
\nomenclature[N]{We}{Weber number}
\nomenclature[N]{Re}{Reynolds number}
\nomenclature[N]{K}{Splashing threshold parameter}
\nomenclature[P]{$\rho$}{Droplet density \nomunit{$[kg/m^3]$}}
\nomenclature[P]{$d_0$}{Droplet diameter before impact \nomunit{$[m]$}}
\nomenclature[P]{$u_0$}{Droplet impact speed \nomunit{$[m/s]$}}
\nomenclature[P]{$\sigma$}{Surface tension \nomunit{$[N/m]$}}
\nomenclature[P]{$\eta$}{Dynamic viscosity \nomunit{$[\Pas]$}}
\nomenclature[P]{$\omega_G$}{Glycerol concentration \nomunit{$[\%]$}}
\nomenclature[P]{$phi$}{Nanoparticle concentration \nomunit{$[\%]$}}
\section{Materials and experimental methods} \label{sec:experiment}
\subsection{Nanofluid preparation}\label{sec:Nano_prep}
Nanofluids with three different mass concentrations ($\phi=$0.01~wt\%, 0.1~wt\% and 1~wt\%) are prepared by diluting a commercial water-based dispersion from Sigma-Aldrich that contains 20~wt\% \Al~nanoparticles with an average diameter of 45~nm (TEM) in aqueous glycerol. The aqueous glycerol solutions are obtained by mixing 99.98\% pure glycerol (Acros Organics) with deionized water. The glycerol concentration $\omega_G$ is then calculated as the mass ratio of the glycerol to the total mass of the nanofluid using an electronic balance with an accuracy of 0.1~mg to weigh the reagents. The final dilution is achieved by mixing via a speed-mixer (SpeedMixer, DAC 150.1 FVZ) and a shaker platform (Heidolph, PROMAX 2020) until well dispersed. The base suspension contains no dispersant, and no surfactant is added during the dilution process. Using this protocol, 30 samples of nanofluids are prepared with glycerol concentrations ranging from 0\% to 80\%. These samples are labeled with the following structure X-$\omega_G$-nf-$\phi$ where X represents the nature of the base liquid, i.e., water (W) or aqueous glycerol (AG).

\subsection{Nanofluid stability}
\begin{figure}[tbp]
         \centering
         \includegraphics[width=0.9\textwidth]{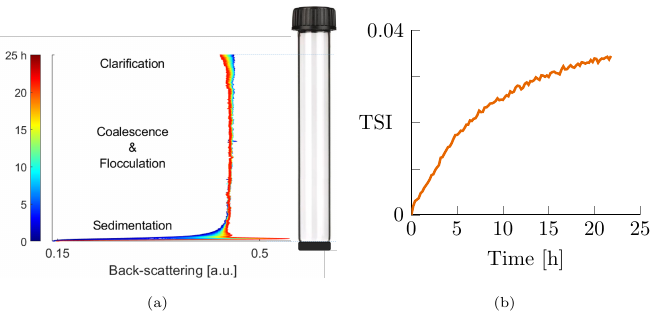}
        \caption{Exemplary stability check of the least stable sample AG-37-nf-1: (a) back-scattering data, (b) $TSI$ over time.}
        \label{fig:turbidity}
\end{figure}

The stability of nanofluids is very important for both reliability and repeatability of the experiments and it strongly depends on the nanoparticle size and concentration~\cite{Vafaei2006}. Although nanofluids are typically considered stable, we confirm the stability of our samples over time using Turbiscan MA2000~\cite{Mengual1999}. Light transmittance is recorded for low mass fraction nanofluids, i.e., $\phi=$0.01~wt\% and 0.1~wt\%, whereas light back-scattering is measured for the higher mass fraction of $\phi=$1~wt\% due to the strong light extinction. By examining the change in transmittance or back-scattering after 25~hours, clarification on top of the tube, coalescence in the middle, and sediment formation at the bottom can be identified~\cite{Mengual1999, Paola2017}. For instance, the destabilization of the least stable sample in our test matrix (AG-37-nf-1) over time is represented in Fig.~\ref{fig:turbidity}a based on the back-scattering data.
In order to quantitatively evaluate the stability over time, the turbidity scan index $TSI$ is defined as~\cite{Li2020, Li2021}:
\begin{equation}
    TSI(t) = \sqrt{\frac{\sum_{i=1}^{n}(x_i^t-x_i^{t=0})^2}{n}},
    \label{eq:TSI}
\end{equation}
where $x_i$ denotes transmittance or back-scattering intensity along the tube length depending on the sample transparency, $n$ is the number of measurement points, and $t$ is the elapsed time. The lower the value of the $TSI$, the better the stability of the nanofluid since increased $TSI$ determines the destabilization. The evolution of the $TSI$ over time for this sample is plotted in Fig.~\ref{fig:TSI}. To be in the safe side in terms of stability, the experimentation time is limited to 4~hours, i.e. $TSI \ll 0.015$. This value corresponds to a change of less than 3\% in the back-scattering along the tube length.

\subsection{Nanofluid characterization}
In addition to stability, the nanofluids are experimentally characterized in terms of density, viscosity and surface tension to accurately calculate the Weber and Reynolds numbers. Viscosity measurements are performed with an ARES-G2 rheometer using a double-walled Couette cell at shear rates between 1 to 100~s$^{-1}$. Extra oscillatory experiments show that while the samples with the highest nanoparticle loading exhibit nonzero storage modulus $G'$, this value is at least an order of magnitude smaller than the loss modulus $G''$ ($\tan \delta = G'' / G' \gg 1$). All in all, the dynamic viscosity is determined within 3\% uncertainty and the nanofluids are accepted to be Newtonian. For the surface tension measurements, the pendant drop method is employed using the KSV CAM200 goniometer. For each sample, the Young-Laplace fit is applied to 150 images: 30 consecutive images out of 5 different drops are averaged to prove that the surface tension does not vary from drop to drop and does not depend on the pending time. Retraction experiments show no wrinkling or other indication that the particles adsorb to the air-liquid interface. Finally, the density of the samples are measured using a pycnometer (BlauBrand, Gay-Lussac pattern, calibrated, 10.109~ml).

The measured properties of these nanofluids are listed in Table~\ref{tab:properties}, in which surface tensions vary from 64 to 72~mN/m and viscosities range from 1 to 70~mPa$\cdot$s. It can be noted that the surface tension values are only slightly affected by the nanoparticles compared to their base liquids. For the fluid properties, there appears a slight non-monotonic behavior in nanofluids compared to their base liquid for some cases, which stays within the uncertainty limits and may occur due to the variations both in the glycerol fraction and the ambient temperature.
\begin{table}[htbp]
\caption{Material properties: mass fraction of glycerol in water $\omega_{G}$, mass fraction of nanoparticles $\phi$, surface tension $\sigma$, density $\rho$ and dynamic viscosity $\eta$.}
\label{tab:properties}
\centering
\begin{tabular}{cccccc} \toprule
\multirow{2}{*}{Sample} & $\omega_{G}$ & $\phi$ & $\sigma$ & $\rho$ & $\eta$ \\ \cmidrule(lr){2-6}
& [wt\%] & [wt\%] & [mN/m] & [kg/m$^3$] & [mPa$\cdot$s] \\ \midrule
        W     &  0  & 0 & 71.29 &  996.83 &  1.05 \\
        AG-30 & 30  & 0 & 69.94 & 1071.38 &  2.51 \\
        AG-37 & 37  & 0 & 68.55 & 1090.20 &  3.28 \\
        AG-43 & 43  & 0 & 68.28 & 1105.25 &  4.14 \\
        AG-58 & 58  & 0 & 66.95 & 1143.92 &  8.69 \\
        AG-70 & 70  & 0 & 66.48 & 1177.35 & 19.69 \\
        AG-75 & 75  & 0 & 65.35 & 1192.80 & 36.61 \\ \addlinespace[0.5em]
    W-nf-0.01 &  0 & 0.01 & 70.22 &  996.71 &  1.03 \\
AG-30-nf-0.01 & 30 & 0.01 & 69.89 & 1070.61 &  2.52 \\
AG-37-nf-0.01 & 37 & 0.01 & 68.49 & 1087.87 &  3.40 \\
AG-43-nf-0.01 & 43 & 0.01 & 67.58 & 1103.08 &  4.19 \\
AG-58-nf-0.01 & 58 & 0.01 & 67.34 & 1145.45 &  9.89 \\
AG-70-nf-0.01 & 70 & 0.01 & 66.12 & 1178.00 & 23.37 \\
AG-75-nf-0.01 & 75 & 0.01 & 65.54 & 1190.22 & 36.29 \\
AG-80-nf-0.01 & 80 & 0.01 & 64.67 & 1204.54 & 60.57 \\ \addlinespace[0.5em]
W-nf-0.1      &  0 & 0.1  & 71.43 &  995.87 &  1.03 \\
AG-30-nf-0.1  & 30 & 0.1  & 69.95 & 1071.05 &  2.62 \\
AG-37-nf-0.1  & 37 & 0.1  & 69.14 & 1086.82 &  3.50 \\
AG-43-nf-0.1  & 43 & 0.1  & 68.31 & 1101.44 &  4.23 \\
AG-58-nf-0.1  & 58 & 0.1  & 67.17 & 1144.09 &  9.19 \\
AG-70-nf-0.1  & 70 & 0.1  & 65.91 & 1178.15 & 23.29 \\
AG-75-nf-0.1  & 75 & 0.1  & 65.04 & 1190.64 & 35.60 \\
AG-80-nf-0.1  & 80 & 0.1  & 64.75 & 1204.31 & 61.98 \\ \addlinespace[0.5em]
AG-30-nf-1    & 30 & 1    & 69.20 & 1080.02 &  2.63 \\
AG-37-nf-1    & 37 & 1    & 69.31 & 1097.64 &  3.46 \\
AG-43-nf-1    & 43 & 1    & 67.85 & 1115.34 &  4.60 \\
AG-58-nf-1    & 58 & 1    & 66.73 & 1154.10 & 10.79 \\
AG-70-nf-1    & 70 & 1    & 65.41 & 1184.32 & 24.63 \\
AG-75-nf-1    & 75 & 1    & 65.09 & 1200.05 & 40.86 \\
AG-80-nf-1    & 80 & 1    & 64.82 & 1214.75 & 69.60 \\ \bottomrule
\end{tabular}
\end{table}
To evaluate the accuracy of the measurements, the current results of the glycerol solutions are compared with literature~\cite{Volk2018, Takamura2012, Cheng2008} and show a maximal difference for viscosity, surface tension, and density being smaller than 6\%, 3\% and 0.1\%, respectively.

The uncertainty analysis for the non-dimensional parameters is carried out according to the procedure detailed by Moffat~\cite{Moffat1988}. The main parameters contributing to uncertainties in $\We$ and $\Re$ numbers come from the nanofluid viscosity and the image post-processing. Indeed, as described below, the droplet diameters and their impact velocities are calculated based on the high-speed camera images. The uncertainty values are designated with $\delta$ in front of each parameter and they are given in Table~\ref{tab:uncertainty}.
\begin{table}[tbp]
    \centering
    \caption{Experimental uncertainties on the governing parameters and non-dimensional groups.}
    \begin{tabular}{ccccc}
        \toprule
        $\delta \eta$ & $\delta d_o$ & $\delta u_o$ & $\delta \Re$ & $\delta \We$ \\ 
        $[\%]$        & $[\%]$       & $[\%]$       & $[\%]$       & $[\%]$ \\ \midrule
         3            & 0.2          & 0.8          & 3.4          & 2.1    \\ \bottomrule
    \end{tabular}
    \label{tab:uncertainty}
\end{table}

\subsection{Experimental setup and conditions}
\begin{figure}[tb]
    \begin {center}
        \includegraphics[scale=0.4]{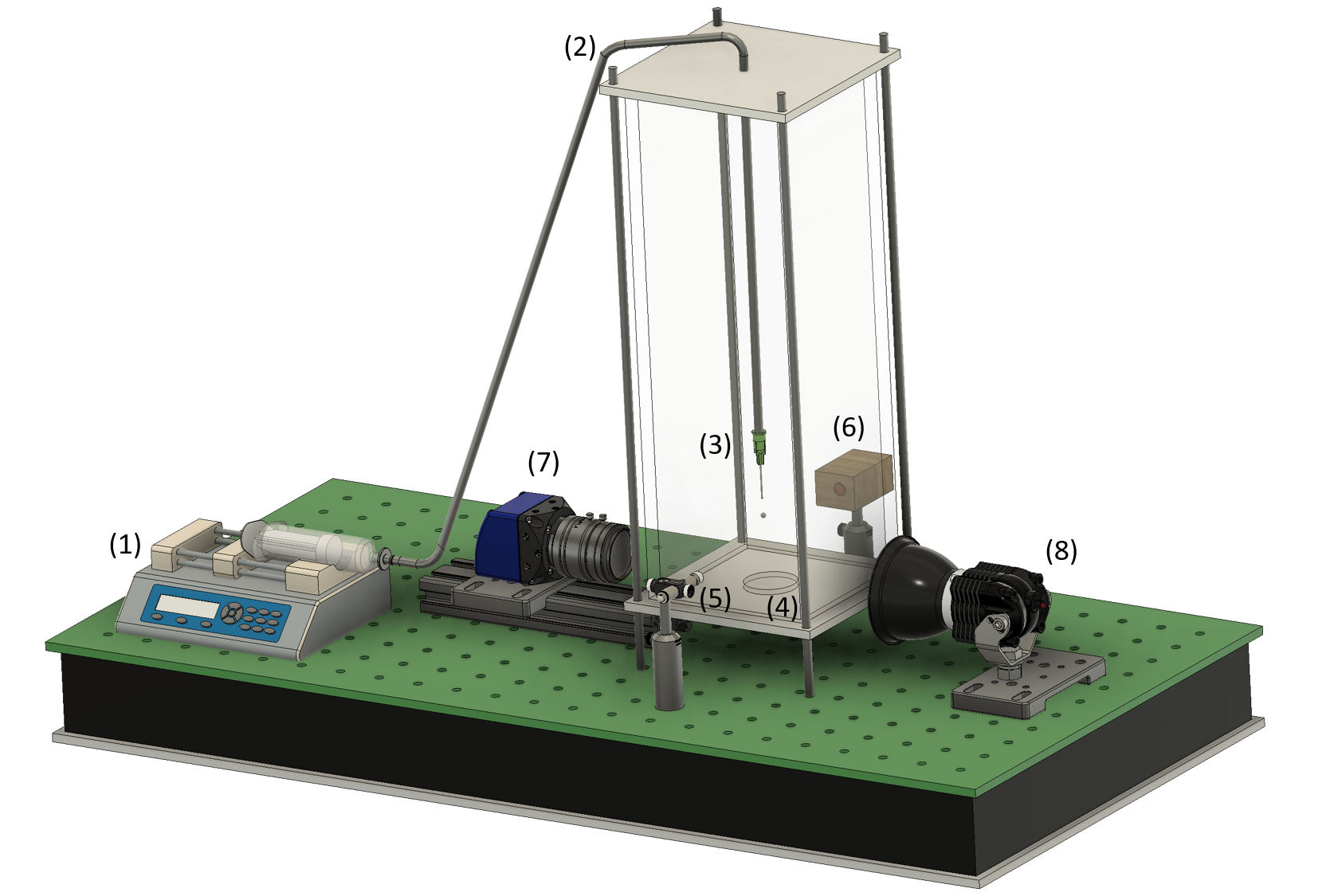}
        \caption{Schematic of the experimental setup characterizing the droplet dynamics: syringe pump (1), fluidic tube (2), blunt-tip needle (3), sapphire substrate (4), laser-diode (5), photo-diode (6), high-speed camera (7), and LED illumination (8).}
        \label{fig:setup}
    \end {center}
\end{figure}

Fig.~\ref{fig:setup} presents a schematic of the experimental setup. A syringe pump (Harvard 4400 PHD) (1) infuses the working fluid through a tube (2) up to a blunt-tip needle (3) where a droplet is formed. To vary the droplet size from 2.2 to 2.8~mm, two needles, i.e, GA 22 and GA 30, are used. The droplet falls onto a 3-cm diameter smooth sapphire substrate (4). The droplet intercepts the beam of a laser-photo-diode unit (5, 6) used to trigger the image acquisition unit. This one is composed of a high-speed camera (Photron Fast-CAM) (7) and a Light-emitting diode (LED) illumination (8). The images are acquired at 12,000~fps in 10-bit gray-scale and with 29~$\mu$s shutter time. Due to the nanoparticles' hazardousness, the drops are enclosed in a transparent Plexiglas box. The roughness of the sapphire substrate is measured by Mitutoyo CS-3200, and the average roughness $R_a$ is found to be 9.09~nm with the roughness pitch of 0.1~nm. They are performed according to ISO 1997 for two different sampling lengths, namely 0.08 and 0.025~mm and each measurement is repeated twice at a different location.

To obtain the desired range of $\We$ and $\Re$ numbers, the drop impact velocity is varied by changing the release height from 10 to 50~cm. The drop diameter and the impact velocity are calculated via image post-processing. In particular, the impact speed of each drop is obtained by analyzing the positions in subsequent frames using a simplified particle tracking velocimetry algorithm. The analyzed images possess a field of view of 896$\times$288 pixels and their typical resolution is about 60~pixels/mm. The drops always have a spherical shape upon impact thanks to the constant volume infusion rate of the syringe pump, allowing us to disregard the shape effect. Their circularity is also cross-checked during the image post-processing. The drops are released onto the same sapphire substrate to exclude the surface roughness effect as well~\cite{Palacios2012}. The substrate is cleaned by an optical cleaning tissue after every drop. Moreover, all experiments are performed at a constant ambient temperature ($21 \pm 1$\deg) and atmospheric ambient pressure. In other words, to keep the effects of nanoparticles on splashing in the focus of attention, other triggering factors, such as surrounding gas pressure or surface structure are excluded from the set of parameters by conducting experiments at atmospheric pressure and always using the same surface.

Excessively long time intervals between consecutive drop impact tests may alter the glycerol concentration at the needle tip due to air humidity. This may be one of the possible reasons of the discrepancies in literature for the splashing thresholds~\cite{Palacios2012}. Therefore, 2-3 drops are always released from the needle tip before each measurement to avoid any variation in the glycerol concentration.
%
\section{Results and discussion}\label{sec:result}
\begin{figure}
    \centering
        \includegraphics[width=0.9\textwidth]{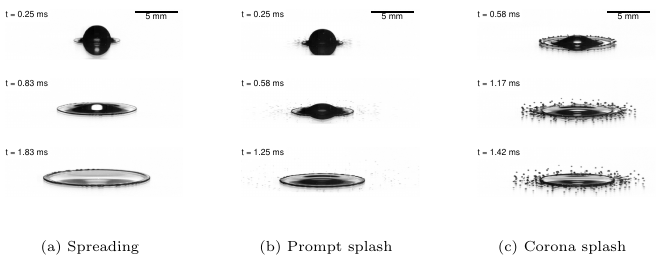}
    \caption{Exemplary droplet impact onto smooth sapphire substrate: (a) Spreading (water, $u_0= 3.46$ m/s), (b) Prompt splash (AG-30-nf-0.1, $u_0= 3.45$ m/s), and (c) Corona splash (AG-70, $u_0= 3.46$ m/s). For better visualization, the background is subtracted.}
    \label{fig:camera_images}
\end{figure}

The results presented in this work are based on 1799 data points from 30 samples. Every impact condition is repeated at least 3 times to ensure repeatability. During the experiments, spreading, prompt splashing, and corona splashing are observed. Example images for these conditions are pictured in Fig.~\ref{fig:camera_images}, and the background is subtracted from the images for better visualization.
In order to evaluate the spreading-to-splashing transition, no distinction is made between prompt and corona splashing~\cite{Stevens2014, VanderWal2006, Aboud2015}.
\subsection{Drop impact outcome regimes}
\begin{figure}
    \centering
         \includegraphics[width=0.9\textwidth]{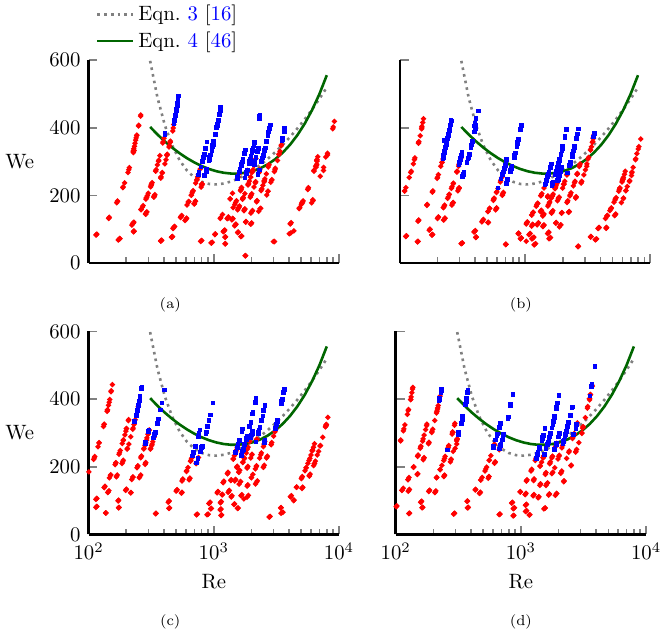}
    \caption{We-Re maps for aqueous solutions of glycerol with (a) 0 wt\% Al$_2$O$_3$ nanofluid, (b) 0.01 wt\% Al$_2$O$_3$ nanofluid, (c) 0.1 wt\% Al$_2$O$_3$ nanofluid, (d) 1 wt\% Al$_2$O$_3$ nanofluid. Red dots represent spreading whereas blue squares stand for splashing. Eqn.~\ref{eq:Palacios} as gray dotted lines and Eqn.~\ref{eq:Amirfazli} is plotted as green solid lines.}
    \label{fig:experiment}
\end{figure}

To determine the spreading-to-splashing transition, the drop impact regimes are plotted on the We-Re maps in Fig.~\ref{fig:experiment}.
Each inclined data set corresponds to a series of impact tests, in which only the impact speeds differ and the rest of the properties ($\rho$, $\eta$, $\sigma$, and $d_0$) remain constant, i.e., constant Ohnesorge number, since it is velocity invariant. Experimental data on the regime maps are marked either as red circles if the drop spreads or as blue squares if splashing occurs, forming spreading and splashing regions. Splashing seems to be totally suppressed at very high or very low $\Re$ numbers.

The experimental data without nanoparticles ($\phi=0$) are illustrated in Fig.~\ref{fig:experiment}a. The correlations reported by Palacios et al.~\cite{Palacios2013}:
\begin{equation}
    \We_\mathrm{crit.} = 5.8 \hspace{3pt} \Re^{0.5} + 4.01\times10^7 \hspace{3pt} \Re^{-1.97},
    \label{eq:Palacios}
\end{equation}
and by Almohammadi and Amirfazli~\cite{Almohammadi2019}:
\begin{equation}
    \We_\mathrm{crit.} = 0.06 \hspace{3pt} \Re + 6762 \hspace{3pt} \Re^{-0.5},
    \label{eq:Amirfazli}
\end{equation}
are also drawn in this figure, which are in good agreement with our experimental data. Since the relation in Eqn.~\ref{eq:Amirfazli} provides a more memorable relationship, its exponents will be used in the subsequent discussion. In fact, the difference between the goodness of fit of both correlations is not larger than 0.05 in terms of $R^2$.

The data for nanofluids for $\phi=$0.01~wt\%, 0.1~wt\%, and 1~wt\% are shown in Fig.~\ref{fig:experiment}b~to~\ref{fig:experiment}d, respectively. The addition of nanoparticles seems to alter $\We_\mathrm{crit.}$ at all $\Re$ numbers. However, this shift is more pronounced at low $\Re$ where we observe an earlier inception of splashing for the nanofluids, i.e., the splashing occurs at much lower $\We_\mathrm{crit.}$ for the nanofluids compared to their base liquid.
\subsection{Lamella spreading velocity measurements}
\begin{figure}
    \centering
        \includegraphics[width=0.8\textwidth]{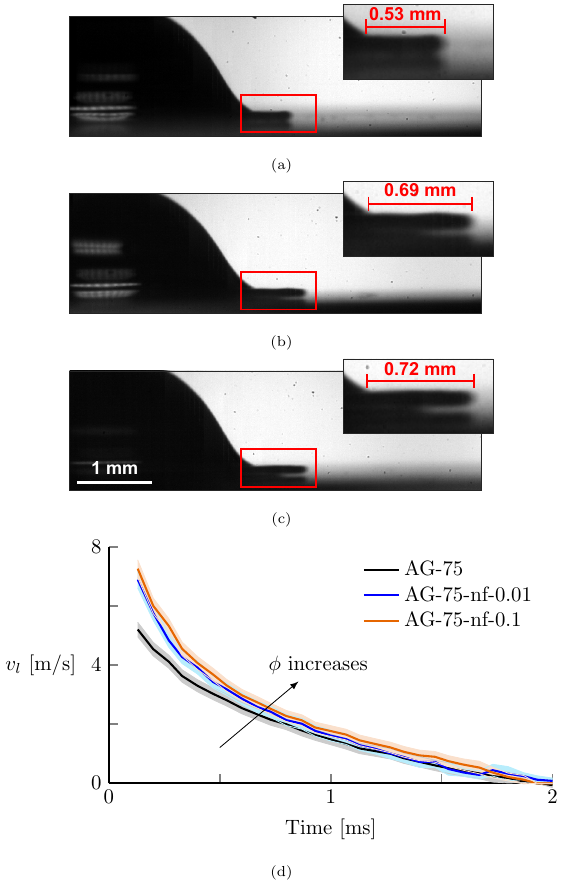}
    \caption{Droplet profiles taken $0.4\pm0.04$ ms after the impact ($u_0 = 2.64 \pm 0.02$~m/s, $\We = 346 \pm 7$, and $\Re = 232 \pm 8$) for (a) AG-75, (b) AG-75-nf-0.01, (c) AG-75-nf-0.1. Earlier lifting of the lamella tip is observed with increased nanoparticle concentration. The scale is the same for all images and inset shows the area highlighted by the red box. For better visualization, the contrast is increased. (d) Lamella spreading speed as a function of time for these samples. The lamella speed is calculated by subtracting the endmost point of the lamella boundary from the previous frame. Therefore, the error bars (shaded areas) are based on the image pixel resolution and the frame rate of the high-speed camera.}
    \label{fig:lamella_lift}
\end{figure}

To understand this earlier inception of the splashing in the low $\Re$ region, we measure the lamella spreading speed $v_l$ in the direction parallel to the substrate. For these experiments, we replace the lens with a long distance microscope (Model K2 DistaMax), with enhanced magnification of 124~pixels/mm. The time resolution is also adapted to 15,000~fps. Fig.~\ref{fig:lamella_lift} displays lamella formation images of three samples obtained at the same $\We$ and $\Re$ numbers ($\We = 346 \pm 7$ and $\Re =232 \pm 8$), for the AG solution and two nanoparticle concentrations ($\phi=$0.01~wt\% and 0.1~wt\%). The three images correspond to the same instant, i.e., $t=0.40\pm 0.04$~ms and are taken at the same location relative to the droplet center.

For this test case, the AG sample undergoes spreading, while the nanofluid samples produce corona splashing. The lamella formed in the presence of nanoparticles is lifted compared to its base liquid. The earlier lifting of the lamella is the consequence of an anticipated wetting~\cite{Mandre2012}. We speculate that this is possibly due to the decreased contact angle with nanoparticle concentration (see Fig.~S2)~\cite{Quetzeri2019n}. That is, the earlier wetting generates a larger momentum in the bulk liquid, which propels the fluid into the lamella. This higher liquid momentum is consistent with the higher $v_l$ during the early spreading stage for the nanofluids, which is illustrated in Fig.~\ref{fig:lamella_lift}d. All of the speeds decrease as the droplets spread, reaching almost identical values after $\approx 1$~ms. No significant difference in the $v_l$ is observed between the two nanoparticle concentrations.

The earlier wetting of the substrate and the formation of a faster lamella have the following consequence. First, an increased $v_l$ induces a larger stress $\Sigma_G$ in the ultra-thin layer of gas beneath the lamella ($\Sigma_G \propto v_l$)~\cite{Liu2015}. Therefore, following the K-H instability model, an increased stress augments the growth rate of the most dangerous mode that grows the fastest $|\omega_m|\propto\Sigma_G^{3/2}$, hence promoting splashing.
\subsection{Spreading-to-splashing transition correlation}
The mathematical correlation for the $\We_\mathrm{crit.}$ expressed in Eqn.~\ref{eq:splash} considers the spreading-to-splashing transition depending only on the $\We$ and $\Re$ numbers. Although the presence of nanoparticles does not modify the bulk properties (e.g., the viscosity and surface tension), it can still alter the interaction of the drop with the substrate. Therefore, a change in the splashing threshold due to presence of nanoparticles in the bulk fluid would not be correctly predicted. Consequently, we propose to include this phenomenon in the mathematical expression via a generalization of the coefficients $K_1$ and $K_2$ as $K_1^{\textup{nf}}(\phi)$ and $K_2^{\textup{nf}}(\phi)$, respectively.

In order to estimate how $K_1^{\textup{nf}}$ and $K_2^{\textup{nf}}$ vary with $\phi$, a least squared curve fitting is implemented to the experimental data, in which these coefficients are taken as fitting parameters at each $\phi$. The exponents are kept constant at $\beta_1=1$ and $\beta_2=0.5$ and splashing threshold is determined as the average value between two consecutive data points, indicating spreading and splashing on a constant $\Oh$ line. The computed fitting parameters are listed in Table~\ref{tab:Knf} for all nanoparticle concentrations. As anticipated, $K_1$ and $K_2$ are in good agreement with literature for the samples without nanoparticles with discrepancies within 0.1\% and 2.4\%, respectively.
\begin{table}[htb]
    \caption{The values of the coefficients $K_1$ and $K_2$ in Eqn.~\ref{eq:splash} ($\beta_1=1$ and $\beta_2=0.5$).}
    \centering
    \begin{tabular}{lcc} \toprule
        Samples                          & $K_1$   & $K_2$\\ \midrule
        Aqueous glycerol solution        & 0.06006 & 6928 \\ \midrule
             & $K_1^{\textup{nf}}$   & $K_2^{\textup{nf}}$\\ \midrule
        0.01~wt\% \Al~nanofluid  & 0.07148 & 4804 \\
        0.1~wt\% \Al~nanofluid   & 0.07633 & 4610 \\
        1~wt\% \Al~nanofluid     & 0.08324 & 4398 \\ \bottomrule
    \end{tabular}
    \label{tab:Knf}
\end{table}
\begin{figure}[htbp]
    \centering
          \includegraphics[width=0.8\textwidth]{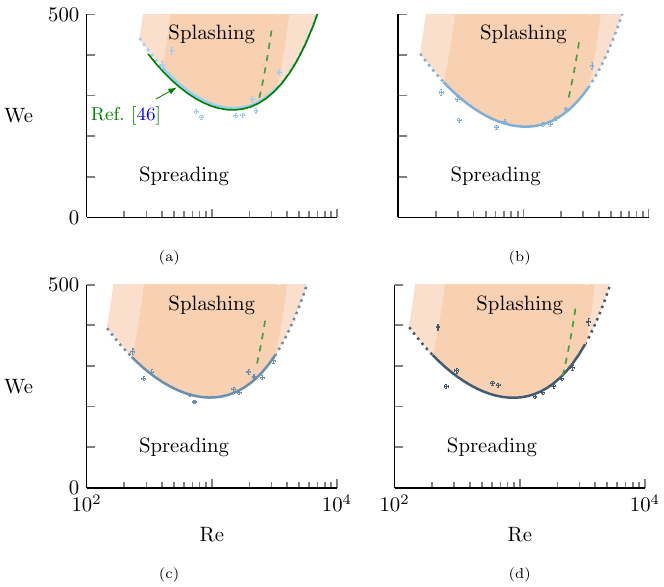}
    \caption{The We-Re maps for aqueous glycerol solutions including (a) 0 wt\% Al$_2$O$_3$ nanofluid, (b) 0.01 wt\% Al$_2$O$_3$ nanofluid, (c) 0.1 wt\% Al$_2$O$_3$ nanofluid, (d) 1 wt\% Al$_2$O$_3$ nanofluid. Within the orange regions, the splashing phenomenon takes place. As the left and the right splashing boundaries are not detected with a high degree of certainty, those areas are designated lighter within their uncertainty limits. Eqn.~\ref{eq:Amirfazli} is plotted as green solid line in (a). The other solid lines represent the fitting curves of Eqn.~\ref{eq:splash} with the coefficients given in Table~\ref{tab:Knf} to the experimental data. The green dashed lines are drawn as a reference ($\Oh=0.0072$) for the prompt-corona transition regions.}
    \label{fig:maps}
\end{figure}

The resultant correlations are plotted in Fig.~\ref{fig:maps}, which demonstrates how the presence of nanoparticles within the fluid modifies the splashing regions. That is, the splashing border on the $\We$-$\Re$ map significantly deviates from Fig.~\ref{fig:maps}a~to~\ref{fig:maps}d due to nanoparticles. In particular, the splashing region becomes broader towards lower $\Re$ numbers. The left and the right boundaries of the splashing regions are determined within uncertainty limits (lighter orange color). Therefore, one should pay attention to the defined ranges of these correlations. The line at $\Oh=0.0072$ is drawn as a reference to approximate the boundary between the prompt and the corona splashing regions qualitatively based on our experimental observations, but a more comprehensive study is necessary to come to a vast conclusion~\cite{Burzynski2020}. 

\begin{figure}[htbp]
    \centering
        \includegraphics[width=0.8\textwidth]{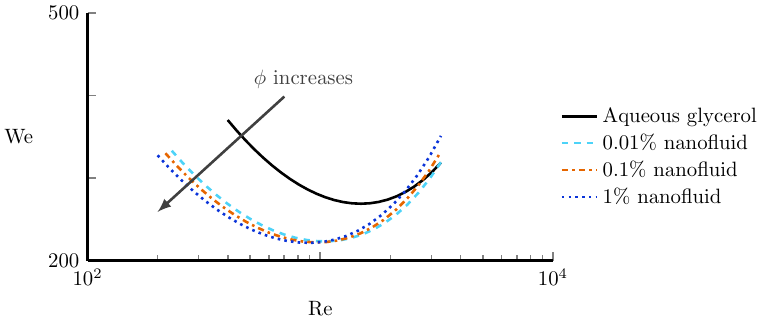}
    \caption{The resultant splashing threshold curves are plotted to clarify the effect of increasing mass fraction of nanoparticles in the nanofluid.}
    \label{fig:curves}
\end{figure}

The splashing transition curves for each nanoparticle concentration are plotted together in Fig.~\ref{fig:curves}.
They exhibit a behavior similar to what is observed for the lamella spreading velocity. The dependence of the splashing threshold on the nanoparticle concentration is non-linear: a low nanoparticle concentration is sufficient to trigger the earlier splashing occurrence whereas with increasing mass fractions, the curves begin to overlap.
\subsection{Generalized spreading-to-splashing transition correlation} 
To represent the observed non-linear dependency, we express the coefficients $K_1^{\textup{nf}}$ and $K_2^{\textup{nf}}$ as functions of $\phi$ in terms of fraction, with the following relation:
\begin{equation}
    K_i^{\textup{nf}}(\phi) = K_i \left (1 + A_i \phi ^{B_i}  \right ),
    \label{eq:Knf}
\end{equation}
for $i=[1,2]$, which comprises four additional fitting parameters. Using a least squared curve fitting procedure, the parameters in Eqn.~\ref{eq:Knf} are computed and tabulated in Table~\ref{tab:Kcorr}.
\begin{table}[htb]
    \caption{Fitting parameters in Eqn.~\ref{eq:Knf}.}
    \centering
    \begin{tabular}{cccc}\toprule
        $A_1$  & $B_1$  & $A_2$   & $B_2$\\ \midrule
        0.7834 & 0.1532 & -0.4350 & 0.038\\ \bottomrule
    \end{tabular}
    \label{tab:Kcorr}
\end{table}

\begin{figure}[htbp]
    \centering
        \includegraphics[width=0.8\textwidth]{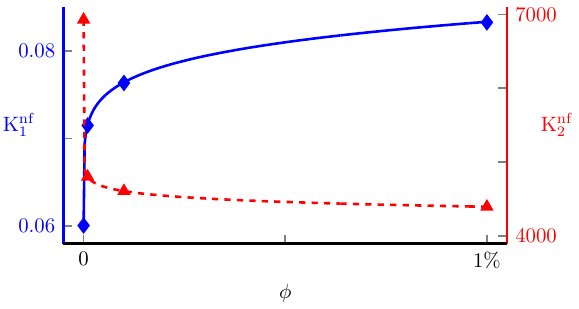}
    \caption{Experimentally found $K_1^{\textup{nf}}$ and $K_2^{\textup{nf}}$ coefficients from Table~\ref{tab:Knf} with respect to nanofluid concentrations are plotted by blue diamonds and red triangles, respectively. The blue solid and the red dashed lines express the correlation in Eqn.~\ref{eq:Knf} with the coefficients in Table~\ref{tab:Kcorr}.}
    \label{fig:correlation}
\end{figure}

In Fig.~\ref{fig:correlation},
the correlations in Eqn.~\ref{eq:Knf} are plotted together with the experimentally found $K_1^{\textup{nf}}$ and $K_2^{\textup{nf}}$ coefficients. Here, the diminishing effect of the increasing nanoparticle concentration is illustrated. Both $K_1^{\textup{nf}}$ and $K_2^{\textup{nf}}$ rapidly reach a plateau value, as predicted by the small exponents $B_1$ and $B_2$. Finally, by substitution of Eqn.~\ref{eq:Knf} into Eqn.~\ref{eq:splash}, we end up with an empirical generalized correlation that incorporates both $\Re$ and $\phi$:
\begin{equation}
    \We_\mathrm{crit.} = K_1 \left (1 + A_1 \phi ^{B_1}  \right ) \Re^{\beta_1} + K_2 \left (1 + A_2 \phi ^{B_2}  \right ) \Re^{-\beta_2}.
    \label{eq:update}
\end{equation}

The critical $\We$ number in Eqn.~\ref{eq:update} for the currently investigated \Al~nanofluid can be considered as the sum of two terms. The first one is only related to the base liquid $\We_\mathrm{crit.,~base} = K_1 \Re^{\beta_1} + K_2 \Re^{-\beta_2}$, while the second one represents the impact of the presence of nanoparticles $\We_\mathrm{crit.,~np} = K_1 A_1 \phi ^{B_1} \Re^{\beta_1} + K_2 A_2 \phi ^{B_2} \Re^{-\beta_2}$. This way of formulating the $\We_\mathrm{crit.}$ is useful to compare the influence of other nanoparticle properties (type, size, aspect ratio, etc.) with respect to the behavior of their base liquid.

The generalized correlation (Eqn.~\ref{eq:update}) is employed for the experimental data of Thoraval et al.~\cite{Thoraval2021} as represented in Fig.~S3. As clearly be seen in that figure, our generalized correlation successfully separates their spreading and splashing points for a higher nanoparticle concentration. Therefore, we can conclude that our suggested correlation has a great potential to be valid for different nanofluid types with various particle sizes even at high concentrations.
%
\section{Conclusions}\label{sec:conclusion}
This study focuses on the experimental investigation on the spreading-to-splashing transition of nanofluid droplets impacted on a smooth surface. The results are represented on several We-Re maps for three nanoparticle concentrations of $\phi~=$~0.01~wt\%, 0.1~wt\%, and 1~wt\%. We cover a large range of $\We$ and $\Re$ numbers, based on droplet diameter, impact speed, and measured material properties, such as the density, viscosity, and surface tension. Moreover, we ensure nanofluid stability during the experimentation time via turbidity measurements. We see that, for pure aqueous glycerol samples, an increase in the viscosity promotes the splashing up to a critical $\Re$, whereupon a further increase demotes the splashing due to the competition between the surrounding gas and the surface tension~\cite{Palacios2013, Mandre2012, Riboux2014, Xu2005, Bird2009}. A similar trend is also present for the nanofluids. This non-monotonic behavior is captured by a modified form of the splashing correlation where the critical $\We$ is shifted for the nanofluids. In fact, nanofluids drastically lower the splashing threshold for smaller $\Re$ numbers. Furthermore, the region where splashing is entirely suppressed shrinks from $\Re < 300$ down to $\Re < 150-200$, depending on the nanoparticle concentration.

For a deeper investigation, lamella spreading speeds of the droplets are recorded with higher resolution and frame rate. During the early spreading stage, the lamella of the nanofluid droplets is noticeably faster than that of their base liquid. Additionally, earlier lifting of the lamella is observed for nanofluid droplets, and we expect this change to be the consequence of early wetting due to the presence of nanoparticles. These findings direct us to Kelvin-Helmholtz instabilities, but to further elucidate the mechanisms underlying the shift in the splashing transition, further research is required. The size and the shape of the lamella should be measured beyond the resolution of the present study.

In order to predict the splash, an empirical correlation is proposed to include both the non-monotonic behavior due to viscosity and nanoparticle concentration in a single expression that covers the previous experimental data on fluids without nanoparticles. In addition, this correlation also fits well on the recently published splashing data of 10-nm silver nanofluids. Interestingly, a modification of the splashing transition occurs even for the lowest nanoparticle concentration of 0.01~wt\%, with the curves beginning to overlap at increased concentrations. Thus, it appears that it is the presence of the nanoparticles, i.e., not their concentration, that adjusts the splashing transition.

This work should also be further expanded to larger (including non-Brownian) and differently shaped particles to determine the limitation of the proposed correlation, since it may depend on the type and the size of the nanoparticles. These results highlight a possible pitfall for using nanoparticles as tracers to track the splashing in ``pure'' fluids. As we prove here, even a small amount of added nanoparticles ($\phi\geq0.01$~wt\%) has an impact on the splashing transition, and the assumption that systems with tracer particles behave identically to those without is shown to be invalid.

Since the nanoparticle stability could potentially have a large influence on the splashing, other methods besides turbidity -- such as Light Extinction Spectroscopy~\cite{Eneren2021} -- can be used to evaluate particle concentration and to monitor agglomerations. As agglomerations are inevitable in many nanoparticle systems due to van der Waals and other colloidal interactions between the small particles, the impact of agglomeration should be tested before industrial use in heat transfer applications.

\section{Acknowledgment}
The authors gratefully acknowledge the in-kind support of Department of Chemical Engineering, KU Leuven. This work was supported by Interne Fondsen KU Leuven / Internal Funds KU Leuven (C24/18/057). The authors also acknowledge Prof. Dr. Pierre Colinet and Prof. Dr. David Seveno for contact angle discussions and measurements, Dr. Anja Vananroye, and Yanshen Zhu for the discussions over nanofluid characterization, and Dr. Balasubramanian Nagarajan for the surface roughness measurements.


\begin{thebibliography}{10}
\expandafter\ifx\csname url\endcsname\relax
  \def\url#1{\texttt{#1}}\fi
\expandafter\ifx\csname urlprefix\endcsname\relax\def\urlprefix{URL }\fi
\expandafter\ifx\csname href\endcsname\relax
  \def\href#1#2{#2} \def\path#1{#1}\fi

\bibitem{Marengo2011}
M.~Marengo, C.~Antonini, I.~V. Roisman, C.~Tropea, Drop collisions with simple
  and complex surfaces, Current Opinion in Colloid {\&} Interface Science
  16~(4) (2011) 292 -- 302.
\newblock \href {https://doi.org/10.1016/j.cocis.2011.06.009}
  {\path{doi:10.1016/j.cocis.2011.06.009}}.

\bibitem{Yarin2006}
A.~Yarin, Drop impact dynamics: Splashing, spreading, receding, bouncing,
  Annual Review of Fluid Mechanics 38~(1) (2006) 159--192.
\newblock \href {https://doi.org/10.1146/annurev.fluid.38.050304.092144}
  {\path{doi:10.1146/annurev.fluid.38.050304.092144}}.

\bibitem{Aksoy2021}
Y.~T. Aksoy, Y.~Zhu, P.~Eneren, E.~Koos, M.~R. Vetrano, The impact of
  nanofluids on droplet/spray cooling of a heated surface: A critical review,
  Energies 14~(1) (2021).
\newblock \href {https://doi.org/10.3390/en14010080}
  {\path{doi:10.3390/en14010080}}.

\bibitem{Minemawari2011}
H.~Minemawari, T.~Yamada, H.~Matsui, J.~Tsutsumi, S.~Haas, R.~Chiba, R.~Kumai,
  T.~Hasegawa, Inkjet printing of single-crystal films, Nature 475~(7356)
  (2011) 364 -- 367.
\newblock \href {https://doi.org/10.1038/nature10313}
  {\path{doi:10.1038/nature10313}}.

\bibitem{Laan2014}
N.~Laan, K.~G. de~Bruin, D.~Bartolo, C.~Josserand, D.~Bonn, Maximum diameter of
  impacting liquid droplets, Phys. Rev. Applied 2 (2014) 044018.
\newblock \href {https://doi.org/10.1103/PhysRevApplied.2.044018}
  {\path{doi:10.1103/PhysRevApplied.2.044018}}.

\bibitem{Josserand2016}
C.~Josserand, S.~Thoroddsen, Drop impact on a solid surface, Annual Review of
  Fluid Mechanics 48~(1) (2016) 365--391.
\newblock \href {https://doi.org/10.1146/annurev-fluid-122414-034401}
  {\path{doi:10.1146/annurev-fluid-122414-034401}}.

\bibitem{Driscoll2010}
M.~M. Driscoll, C.~S. Stevens, S.~R. Nagel, Thin film formation during
  splashing of viscous liquids, Phys. Rev. E 82 (2010) 036302.
\newblock \href {https://doi.org/10.1103/PhysRevE.82.036302}
  {\path{doi:10.1103/PhysRevE.82.036302}}.

\bibitem{Rein1993}
M.~Rein, Phenomena of liquid drop impact on solid and liquid surfaces, Fluid
  Dynamics Research 12~(2) (1993) 61 -- 93.
\newblock \href {https://doi.org/10.1016/0169-5983(93)90106-K}
  {\path{doi:10.1016/0169-5983(93)90106-K}}.

\bibitem{Vega2017}
E.~J. Vega, A.~A. Castrej{\'o}n-Pita, Suppressing prompt splash with polymer
  additives, Experiments in Fluids 58~(5) (2017) 57.
\newblock \href {https://doi.org/10.1007/s00348-017-2341-y}
  {\path{doi:10.1007/s00348-017-2341-y}}.

\bibitem{Rioboo2001}
R.~Rioboo, C.~Tropea, M.~Marengo, Outcomes from a drop impact on solid
  surfaces, Atomization and Sprays 11~(2) (2001) 155--165.
\newblock \href {https://doi.org/10.1615/AtomizSpr.v11.i2.40}
  {\path{doi:10.1615/AtomizSpr.v11.i2.40}}.

\bibitem{Xu2007}
L.~Xu, L.~Barcos, S.~R. Nagel, Splashing of liquids: Interplay of surface
  roughness with surrounding gas, Phys. Rev. E 76 (2007) 066311.
\newblock \href {https://doi.org/10.1103/PhysRevE.76.066311}
  {\path{doi:10.1103/PhysRevE.76.066311}}.

\bibitem{Quetzeri2019}
M.~A. Quetzeri-Santiago, K.~Yokoi, A.~A. Castrej\'on-Pita, J.~R.
  Castrej\'on-Pita, Role of the dynamic contact angle on splashing, Phys. Rev.
  Lett. 122 (2019) 228001.
\newblock \href {https://doi.org/10.1103/PhysRevLett.122.228001}
  {\path{doi:10.1103/PhysRevLett.122.228001}}.

\bibitem{Zhang2017}
B.~Zhang, J.~Li, P.~Guo, Q.~Lv, Experimental studies on the effect of reynolds
  and weber numbers on the impact forces of low-speed droplets colliding with a
  solid surface, Experiments in Fluids 58~(9) (2017) 125.
\newblock \href {https://doi.org/10.1007/s00348-017-2413-z}
  {\path{doi:10.1007/s00348-017-2413-z}}.

\bibitem{Mundo1995}
C.~Mundo, M.~Sommerfeld, C.~Tropea, Droplet-wall collisions: Experimental
  studies of the deformation and breakup process, International Journal of
  Multiphase Flow 21~(2) (1995) 151 -- 173.
\newblock \href {https://doi.org/10.1016/0301-9322(94)00069-V}
  {\path{doi:10.1016/0301-9322(94)00069-V}}.

\bibitem{Stow1981}
C.~D. Stow, M.~G. Hadfield, J.~M. Ziman, An experimental investigation of fluid
  flow resulting from the impact of a water drop with an unyielding dry
  surface, Proceedings of the Royal Society of London. A. Mathematical and
  Physical Sciences 373~(1755) (1981) 419--441.
\newblock \href {https://doi.org/10.1098/rspa.1981.0002}
  {\path{doi:10.1098/rspa.1981.0002}}.

\bibitem{Palacios2013}
J.~Palacios, J.~Hernandez, P.~Gomez, C.~Zanzi, J.~Lopez, Experimental study of
  splashing patterns and the splashing/deposition threshold in drop impacts
  onto dry smooth solid surfaces, Experimental Thermal and Fluid Science 44
  (2013) 571 -- 582.
\newblock \href {https://doi.org/10.1016/j.expthermflusci.2012.08.020}
  {\path{doi:10.1016/j.expthermflusci.2012.08.020}}.

\bibitem{Mandre2012}
S.~Mandre, M.~P. Brenner, The mechanism of a splash on a dry solid surface,
  Journal of Fluid Mechanics 690 (2012) 148–172.
\newblock \href {https://doi.org/10.1017/jfm.2011.415}
  {\path{doi:10.1017/jfm.2011.415}}.

\bibitem{Thoroddsen2010}
S.~T. Thoroddsen, K.~Takehara, T.~G. Etoh, Bubble entrapment through
  topological change, Physics of Fluids 22~(5) (2010) 051701.
\newblock \href {https://doi.org/10.1063/1.3407654}
  {\path{doi:10.1063/1.3407654}}.

\bibitem{Palacios2012}
J.~Palacios, J.~Hern{\'a}ndez, P.~G{\'o}mez, C.~Zanzi, J.~L{\'o}pez, On the
  impact of viscous drops onto dry smooth surfaces, Experiments in Fluids
  52~(6) (2012) 1449--1463.
\newblock \href {https://doi.org/10.1007/s00348-012-1264-x}
  {\path{doi:10.1007/s00348-012-1264-x}}.

\bibitem{Allen1975}
R.~Allen, The role of surface tension in splashing, Journal of Colloid and
  Interface Science 51~(2) (1975) 350 -- 351.
\newblock \href {https://doi.org/10.1016/0021-9797(75)90126-5}
  {\path{doi:10.1016/0021-9797(75)90126-5}}.

\bibitem{Liu2015}
Y.~Liu, P.~Tan, L.~Xu, Kelvin{\textendash}helmholtz instability in an ultrathin
  air film causes drop splashing on smooth surfaces, Proceedings of the
  National Academy of Sciences 112~(11) (2015) 3280--3284.
\newblock \href {https://doi.org/10.1073/pnas.1417718112}
  {\path{doi:10.1073/pnas.1417718112}}.

\bibitem{Kolinski2012}
J.~M. Kolinski, S.~M. Rubinstein, S.~Mandre, M.~P. Brenner, D.~A. Weitz,
  L.~Mahadevan, Skating on a film of air: Drops impacting on a surface, Phys.
  Rev. Lett. 108 (2012) 074503.
\newblock \href {https://doi.org/10.1103/PhysRevLett.108.074503}
  {\path{doi:10.1103/PhysRevLett.108.074503}}.

\bibitem{Burzynski2019}
D.~A. Burzynski, S.~E. Bansmer, Role of surrounding gas in the outcome of
  droplet splashing, Phys. Rev. Fluids 4 (2019) 073601.
\newblock \href {https://doi.org/10.1103/PhysRevFluids.4.073601}
  {\path{doi:10.1103/PhysRevFluids.4.073601}}.

\bibitem{Xu2005}
L.~Xu, W.~W. Zhang, S.~R. Nagel, Drop splashing on a dry smooth surface, Phys.
  Rev. Lett. 94 (2005) 184505.
\newblock \href {https://doi.org/10.1103/PhysRevLett.94.184505}
  {\path{doi:10.1103/PhysRevLett.94.184505}}.

\bibitem{Riboux2014}
G.~Riboux, J.~M. Gordillo, Experiments of drops impacting a smooth solid
  surface: A model of the critical impact speed for drop splashing, Phys. Rev.
  Lett. 113 (2014) 024507.
\newblock \href {https://doi.org/10.1103/PhysRevLett.113.024507}
  {\path{doi:10.1103/PhysRevLett.113.024507}}.

\bibitem{Stevens2014}
C.~S. Stevens, Scaling of the splash threshold for low-viscosity fluids, {EPL}
  (Europhysics Letters) 106~(2) (2014) 24001.
\newblock \href {https://doi.org/10.1209/0295-5075/106/24001}
  {\path{doi:10.1209/0295-5075/106/24001}}.

\bibitem{Roisman2015}
I.~V. Roisman, A.~Lembach, C.~Tropea, Drop splashing induced by target
  roughness and porosity: The size plays no role, Advances in Colloid and
  Interface Science 222 (2015) 615--621, reinhard Miller, Honorary Issue.
\newblock \href {https://doi.org/10.1016/j.cis.2015.02.004}
  {\path{doi:10.1016/j.cis.2015.02.004}}.

\bibitem{Latka2012}
A.~Latka, A.~Strandburg-Peshkin, M.~M. Driscoll, C.~S. Stevens, S.~R. Nagel,
  Creation of prompt and thin-sheet splashing by varying surface roughness or
  increasing air pressure, Phys. Rev. Lett. 109 (2012) 054501.
\newblock \href {https://doi.org/10.1103/PhysRevLett.109.054501}
  {\path{doi:10.1103/PhysRevLett.109.054501}}.

\bibitem{Ersayin2013}
E.~Ersayın, F.~Selimefendigil, Numerical investigation of impinging jets with
  nanofluids on a moving plate, Mathematical and Computational Applications
  18~(3) (2013) 428--437.
\newblock \href {https://doi.org/10.3390/mca18030428}
  {\path{doi:10.3390/mca18030428}}.

\bibitem{Basaran2013}
A.~Başaran, F.~Selimefendigil, Numerical study of heat transfer due to
  twinjets impingement onto an isothermal moving plate, Mathematical and
  Computational Applications 18~(3) (2013) 340--350.
\newblock \href {https://doi.org/10.3390/mca18030340}
  {\path{doi:10.3390/mca18030340}}.

\bibitem{Buonomo2019}
B.~Buonomo, O.~Manca, N.~S. Bondareva, M.~A. Sheremet, Thermal and fluid
  dynamic behaviors of confined slot jets impinging on an isothermal moving
  surface with nanofluids, Energies 12~(11) (2019).
\newblock \href {https://doi.org/10.3390/en12112074}
  {\path{doi:10.3390/en12112074}}.

\bibitem{Figueiredo2020}
M.~Figueiredo, G.~Marseglia, A.~S. Moita, M.~R.~O. Panão, A.~P.~C. Ribeiro,
  C.~M. Medaglia, A.~L.~N. Moreira, Thermofluid characterization of nanofluid
  spray cooling combining phase doppler interferometry with high-speed
  visualization and time-resolved ir thermography, Energies 13~(22) (2020).
\newblock \href {https://doi.org/10.3390/en13225864}
  {\path{doi:10.3390/en13225864}}.

\bibitem{Kang2019}
B.~Kang, M.~Marengo, S.~Begg, A study of the effect of nanoparticle
  concentration on the characteristics of nanofluid sprays, Journal of Applied
  Fluid Mechanics 12 (2019) 413 -- 420.
\newblock \href {https://doi.org/10.29252/jafm.12.02.29182}
  {\path{doi:10.29252/jafm.12.02.29182}}.

\bibitem{Maly2019}
M.~Maly, A.~S. Moita, J.~Jedelsky, A.~P.~C. Ribeiro, A.~L.~N. Moreira, Effect
  of nanoparticles concentration on the characteristics of nanofluid sprays for
  cooling applications, Journal of Thermal Analysis and Calorimetry 135~(6)
  (2019) 3375--3386.
\newblock \href {https://doi.org/10.1007/s10973-018-7444-z}
  {\path{doi:10.1007/s10973-018-7444-z}}.

\bibitem{Vafaei2006}
S.~Vafaei, T.~Borca-Tasciuc, M.~Z. Podowski, A.~Purkayastha, G.~Ramanath, P.~M.
  Ajayan, Effect of nanoparticles on sessile droplet contact angle,
  Nanotechnology 17~(10) (2006) 2523--2527.
\newblock \href {https://doi.org/10.1088/0957-4484/17/10/014}
  {\path{doi:10.1088/0957-4484/17/10/014}}.

\bibitem{Mengual1999}
O.~Mengual, G.~Meunier, I.~Cayré, K.~Puech, P.~Snabre, Turbiscan ma 2000:
  multiple light scattering measurement for concentrated emulsion and
  suspension instability analysis, Talanta 50~(2) (1999) 445--456.
\newblock \href {https://doi.org/10.1016/S0039-9140(99)00129-0}
  {\path{doi:10.1016/S0039-9140(99)00129-0}}.

\bibitem{Paola2017}
M.~G.~D. Paola, V.~Calabr{\`{o}}, M.~D. Simone, Light scattering methods to
  test inorganic {PCMs} for application in buildings, {IOP} Conference Series:
  Materials Science and Engineering 251 (2017) 012122.
\newblock \href {https://doi.org/10.1088/1757-899x/251/1/012122}
  {\path{doi:10.1088/1757-899x/251/1/012122}}.

\bibitem{Li2020}
S.~Li, Y.~H. Ng, H.~C. Lau, O.~Torsæter, L.~P. Stubbs, Experimental
  investigation of stability of silica nanoparticles at reservoir conditions
  for enhanced oil-recovery applications, Nanomaterials 10~(8) (2020).
\newblock \href {https://doi.org/10.3390/nano10081522}
  {\path{doi:10.3390/nano10081522}}.

\bibitem{Li2021}
B.~Li, C.~Zou, H.~Liang, W.~Chen, S.~Lin, Y.~Liao, Mass transfer from nanofluid
  single drops in low interfacial tension liquid–liquid extraction process,
  Chemical Physics Letters 771 (2021) 138530.
\newblock \href {https://doi.org/10.1016/j.cplett.2021.138530}
  {\path{doi:10.1016/j.cplett.2021.138530}}.

\bibitem{Volk2018}
A.~Volk, C.~J. K{\"a}hler, Density model for aqueous glycerol solutions,
  Experiments in Fluids 59~(5) (2018) 75.
\newblock \href {https://doi.org/10.1007/s00348-018-2527-y}
  {\path{doi:10.1007/s00348-018-2527-y}}.

\bibitem{Takamura2012}
K.~Takamura, H.~Fischer, N.~R. Morrow, Physical properties of aqueous glycerol
  solutions, Journal of Petroleum Science and Engineering 98-99 (2012) 50 --
  60.
\newblock \href {https://doi.org/10.1016/j.petrol.2012.09.003}
  {\path{doi:10.1016/j.petrol.2012.09.003}}.

\bibitem{Cheng2008}
N.-S. Cheng, Formula for the viscosity of a glycerol-water mixture, Industrial
  \& Engineering Chemistry Research 47~(9) (2008) 3285--3288.
\newblock \href {https://doi.org/10.1021/ie071349z}
  {\path{doi:10.1021/ie071349z}}.

\bibitem{Moffat1988}
R.~J. Moffat, Describing the uncertainties in experimental results,
  Experimental Thermal and Fluid Science 1~(1) (1988) 3--17.
\newblock \href {https://doi.org/10.1016/0894-1777(88)90043-X}
  {\path{doi:10.1016/0894-1777(88)90043-X}}.

\bibitem{VanderWal2006}
R.~L.~V. Wal, G.~M. Berger, S.~D. Mozes, The splash/non-splash boundary upon a
  dry surface and thin fluid film, Experiments in Fluids 40~(1) (2006) 53--59.
\newblock \href {https://doi.org/10.1007/s00348-005-0045-1}
  {\path{doi:10.1007/s00348-005-0045-1}}.

\bibitem{Aboud2015}
D.~G.~K. Aboud, A.-M. Kietzig, Splashing threshold of oblique droplet impacts
  on surfaces of various wettability, Langmuir 31~(36) (2015) 10100--10111.
\newblock \href {https://doi.org/10.1021/acs.langmuir.5b02447}
  {\path{doi:10.1021/acs.langmuir.5b02447}}.

\bibitem{Almohammadi2019}
H.~Almohammadi, A.~Amirfazli, Droplet impact: Viscosity and wettability effects
  on splashing, Journal of Colloid and Interface Science 553 (2019) 22 -- 30.
\newblock \href {https://doi.org/10.1016/j.jcis.2019.05.101}
  {\path{doi:10.1016/j.jcis.2019.05.101}}.

\bibitem{Quetzeri2019n}
M.~A. Quetzeri-Santiago, A.~A. Castrej{\'o}n-Pita, J.~R. Castrej{\'o}n-Pita,
  The effect of surface roughness on the contact line and splashing dynamics of
  impacting droplets, Scientific Reports 9~(1) (2019) 15030.
\newblock \href {https://doi.org/10.1038/s41598-019-51490-5}
  {\path{doi:10.1038/s41598-019-51490-5}}.

\bibitem{Burzynski2020}
D.~A. Burzynski, I.~V. Roisman, S.~E. Bansmer, On the splashing of high-speed
  drops impacting a dry surface, Journal of Fluid Mechanics 892 (2020) A2.
\newblock \href {https://doi.org/10.1017/jfm.2020.168}
  {\path{doi:10.1017/jfm.2020.168}}.

\bibitem{Thoraval2021}
M.-J. Thoraval, J.~Schubert, S.~Karpitschka, M.~Chanana, F.~Boyer,
  E.~Sandoval-Naval, J.~F. Dijksman, J.~H. Snoeijer, D.~Lohse, Nanoscopic
  interactions of colloidal particles can suppress millimetre drop splashing,
  Soft Matter 17 (2021) 5116--5121.
\newblock \href {https://doi.org/10.1039/D0SM01367F}
  {\path{doi:10.1039/D0SM01367F}}.

\bibitem{Bird2009}
J.~C. Bird, S.~S.~H. Tsai, H.~A. Stone, Inclined to splash: triggering and
  inhibiting a splash with tangential velocity, New Journal of Physics 11~(6)
  (2009) 063017.
\newblock \href {https://doi.org/10.1088/1367-2630/11/6/063017}
  {\path{doi:10.1088/1367-2630/11/6/063017}}.

\bibitem{Eneren2021}
P.~Eneren, Y.~T. Aksoy, Y.~Zhu, E.~Koos, M.~R. Vetrano, Light extinction
  spectroscopy applied to polystyrene colloids: Sensitivity to complex
  refractive index uncertainties and to noise, Journal of Quantitative
  Spectroscopy and Radiative Transfer 261 (2021) 107494.
\newblock \href {https://doi.org/https://doi.org/10.1016/j.jqsrt.2020.107494}
  {\path{doi:https://doi.org/10.1016/j.jqsrt.2020.107494}}.

\end{thebibliography}
\end{document}